\documentclass[preprint]{revtex4-2}

\usepackage{enumerate}
\usepackage{enumitem}
\usepackage{graphicx}
\usepackage{dcolumn}
\usepackage{bm}
\usepackage{graphicx}
\usepackage{amssymb}
\usepackage{epstopdf}
\usepackage{color}
\usepackage{xparse}

\usepackage[english]{babel}
\usepackage{soul}

\begin{document}
\raggedbottom

\title{\Large \textbf{
Stoichiometric control of electron mobility and 2D superconductivity at LaAlO$_3$-SrTiO$_3$ interfaces} }

\author{Gyanendra Singh$^{1*}$, Roger Guzman$^2$, Guilhem Saïz$^3$, Wu Zhou$^2$, Jaume Gazquez$^1$, Jordi Fraxedas$^5$, Fereshteh Masoudinia$^1$, Dag Winkler$^4$, Tord Claeson$^4$, Nicolas Bergeal$^3$, Gervasi Herranz$^1$, Alexei Kalaboukhov$^4$}

\affiliation{
\vskip 0.6cm $^1$ Institut de Ciència de Materials de Barcelona (ICMAB-CSIC), Campus de la UAB, 08193 Bellaterra, Catalonia, Spain.\\
$^2$ School of Physical Sciences, University of Chinese Academy of Sciences Beijing 100049, China.\\
$^3$ Laboratoire de Physique et d’Etude des Materiaux, ESPCI Paris, Universite PSL, CNRS, Sorbonne Universite, Paris, France.\\
$^4$ Department of Microtechnology and Nanoscience - MC2, Chalmers University of Technology, SE 412 96 Gothenburg, Sweden. \\
$^5$ Catalan Institute of Nanoscience and Nanotechnology (ICN2), CSIC and BIST, Campus UAB, Bellaterra, Barcelona 08193, Spain.
}
\date{\today}

\email[]{gsingh@icmab.es}

\maketitle


\textbf{SrTiO$_3$-based conducting interfaces, which exhibit coexistence of gate-tunable 2D superconductivity and strong Rashba spin-orbit coupling (RSOC), are candidates to host topological superconductive phases. Yet, superconductivity is usually in the dirty limit, which tends to suppress nonconventional pairing and therefore challenges these expectations. Here we report on LaAlO$_3$/SrTiO$_3$ (LAO/STO) interfaces with remarkably large mobility and mean free paths comparable to the superconducting coherence length, approaching the clean limit for superconductivity. We further show that the carrier density, mobility, and formation of the superconducting condensate are controlled by the fine-tuning of La/Al chemical ratio in the LAO film. Interestingly, we find a region in the superconducting phase diagram where the critical temperature is not suppressed below the Lifshitz transition, at odds with previous experimental investigations. These findings point out the relevance of achieving a clean-limit regime to enhance the observation of unconventional pairing mechanisms in these systems.
}
\vskip 0.6cm

\textbf{\large Introduction}\\
The coexistence of two-dimensional (2D) superconductivity and spin-orbit-coupling (SOC) is key for the development of novel device concepts for quantum technologies that exploit unconventional electron pairing in topological phases \cite{Elsa2020,Alicea2012}. Boundaries and defects in topological superconductors can host zero-energy Majorana modes, offering applications in quantum computation \cite{Bernevig2013, Masatoshi2019, Manchon2015, Feliciano2020, Mohanta2014,Maria2018}. Although seminal proposals focused on hybrid superconductor-semiconductor structures \cite{Mourik2012, Lutchyn2018}, the simultaneous presence of superconductivity and large tunable SOC in SrTiO$_3$ quasi-two dimensional electron gases (q-2DEGs) offers a particularly appealing platform to explore unconventional and topological superconductive phases \cite{Perroni2019, Barthelemy2021}. Particularly, q-2DEGs at the LaAlO$_3$/SrTiO$_3$ (LAO/STO) interface have been investigated intensively, where electrostatic gating allows to control the electron band filling within the t$_{2g}$ manifold \cite{Gariglio2016,Salluzzo2009,Herranz2015}, enabling the electric field modulation of the superconducting phase diagram \cite{Caviglia2008,Hurand2015,Joshua2012, Singh2019,Singh2022,Scheurer2015,Stornaiuolo2017} and Rashba SOC \cite{Caviglia2010,Shalom2010,Singh2017}. Interestingly, the multiband character of the LAO/STO q-2DEG together with large Rashba SOC promotes the appearance of singlet-triplet mixed pairings, which, in the presence of magnetic fields, could lead to nontrivial superconductive states \cite{Lepori2021}. However, while Anderson's theorem holds for singlet pairing \cite{Anderson1959}, disorder is expected to reduce triplet pairing, potentially hindering the emergence of topological phases. It is therefore indispensable to approach a clean limit for superconductivity at the LAO/STO q-2DEGs to enhance the chances to detect nontrivial topological features in these systems \cite{Jouan2020}. 
\vskip 0.4cm

In this work, we report on LAO/STO q-2DEGs with exceptionally large mobilities, which allow for achieving mean free paths ($l_{mfp}$) comparable to the superconductive coherence length ($\xi$), approaching the clean limit even in the depletion regime. In LAO/STO interfaces the electronic conductivity is ensured by t$_{2g}$ electrons in Ti (d$_{xy}$, d$_{xz}$, d$_{yz}$) orbitals. The quantum confinement of the q-2DEG splits the energy levels of the 3d orbitals, whose splitting is dictated by the out-of-plane effective mass m$^*$, which depends on the degree of orbital wavefunctions overlapping along the confinement direction \cite{Santander2011}. In conventional (001) oriented interfaces, the degenerated d$_{xz/yz}$ orbitals lie at higher energies, and their wave function delocalizes deeper into STO with a dielectric constant similar to that of the bulk. Electrons occupying these subbands encounter less scattering, resulting in higher electronic mobilities. In contrast, electrons in d$_{xy}$ orbitals lie at lower energies and are confined near the interface where the dielectric constant is reduced by the interfacial electric field and shows smaller electronic mobility due to the larger scattering caused by imperfections (see figure 1b,c) \cite{Berner2013,Valentinis2017,Biscaras2012}. The filling of these subbands can be precisely controlled with a gate voltage V$_G$ by continuous and reversible doping of electrons at the quantum well and monitored by Hall effect \cite{Biscaras2012}. The critical temperature T$_c$ shows a dome-shaped phase diagram, first increasing at low doping in the depletion region and then going through a maximum $\mathrm{T}_{c}^{max}$ to subsequently reduce for further increasing V$_G$ towards the overdoped region \cite{Caviglia2008,Hurand2015,Joshua2012}. Attempts have been made to correlate this dome-shaped phase diagram with the filling of the subbands. It has been suggested that the superconducting order parameter emerges when electrons start filling d$_{xz/yz}$ subbands, whereas it is strongly suppressed or even absent below the Lifshitz transition \cite{Herranz2015,Caviglia2008,Joshua2012,Biscaras2012,Singh2018,Singh2017,Nicola2019}. In line with this argument, only one superconducting gap has been reported so far in (001)-oriented interfaces, implying the absence of d$_{xy}$ superconductivity for this orientation \cite{Richter2013}. This has precluded the investigation of the superconductivity of the (001) LAO/STO quantum well in the dilute regime, where one single band is occupied. We stress that in this region of phase diagram, the average electronic mobility for d$_{xy}$ electrons usually reduces to $\mu_{d_{xy}}$ $<$ 50 cm$^2$/V.s, corresponding to a mean free path $l_{mfp}$ $<$ 1nm, which leads these systems to the extreme dirty limit \cite{Herranz2015,Joshua2012,Biscaras2012,Singh2018,Singh2017,Nicola2019}. As we discuss below, we reach mobilities around two orders of magnitude larger for d$_{xy}$ electrons, which is crucial to observe superconductivity mediated by d$_{xy}$ bands. 
\begin{figure*}[t]
\begin{center}
 \includegraphics [width=16cm]{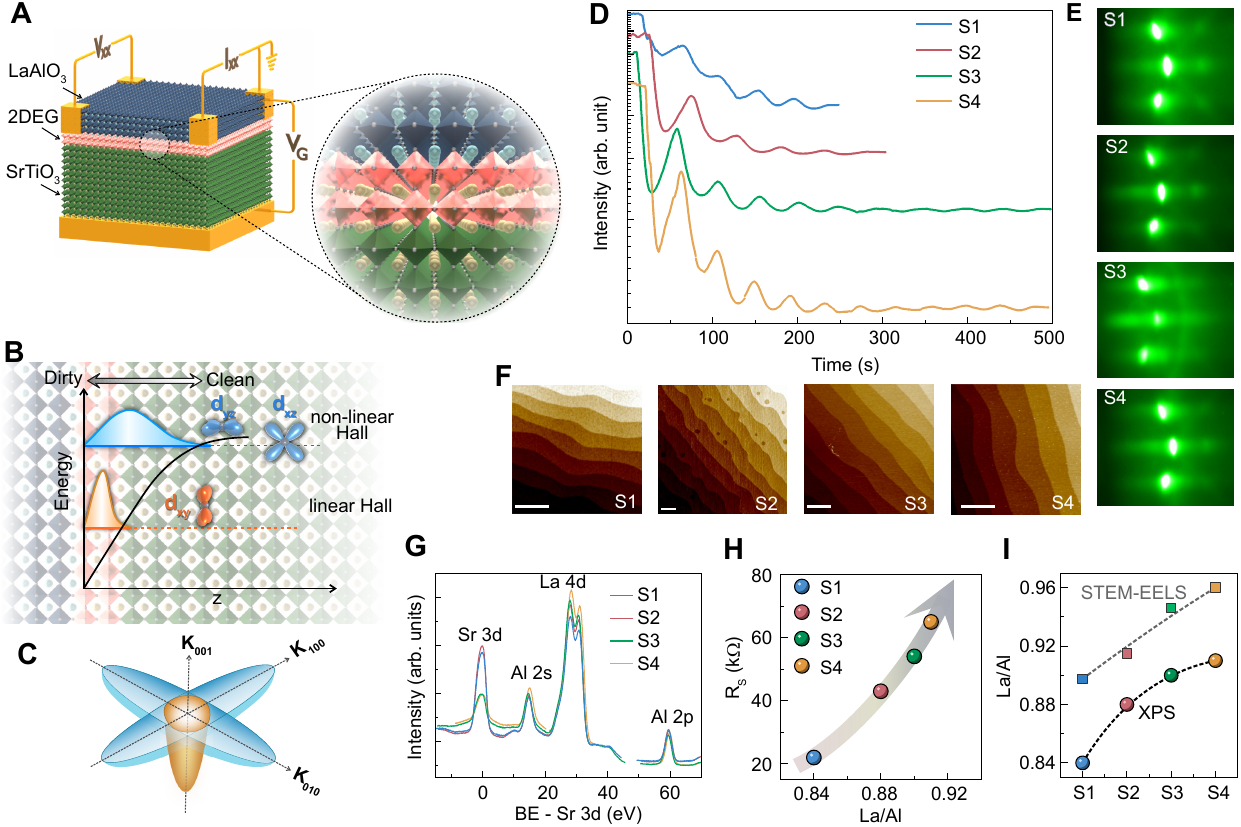}%
\end{center}
\vskip -0.5cm
\caption{\label{fig1} \textbf{a} Schematic illustration of LaAlO$_3$/SrTiO$_3$ heterostructure with two-dimensional electron gas (2DEG) at the interface.  \textbf{b, c} Sketch of the quantum well in real and momentum space, respectively. Three non-degenerate t$_{2g}$ orbitals split due to the confinement potential: the d$_{xz/yz}$ orbitals (blue) are shifted higher in energy with respect to the d$_{xy}$ orbitals (red). The conduction electrons, which correspond to d$_{xy}$ orbitals, are closer to the interface, showing lower electron mobility compared to the mobility for d$_{xz/yz}$ orbitals, which are extended deeper into the STO. \textbf{d} RHEED intensity as a function of time, depicting a layer-by-layer growth, from which thickness can be inferred for the LaAlO$_3$ layer. \textbf{e} RHEED patterns obtained during the PLD growth show a smooth film deposition. \textbf{f} Atomic force microscopy (AFM) topography images of a film. The well-aligned terraces indicate the surface smoothness. Scale bar: 500 nm. \textbf{g} The XPS spectra for all films. \textbf{h} The modulation of sheet resistance at room temperature as a function of the ratio of La/Al extracted from XPS spectra. \textbf{i} La/Al ratio in different samples extracted from STEM-EELS and XPS analysis.}
\end{figure*}

\begin{figure*}[t]
\begin{center}
\includegraphics [width=16cm]{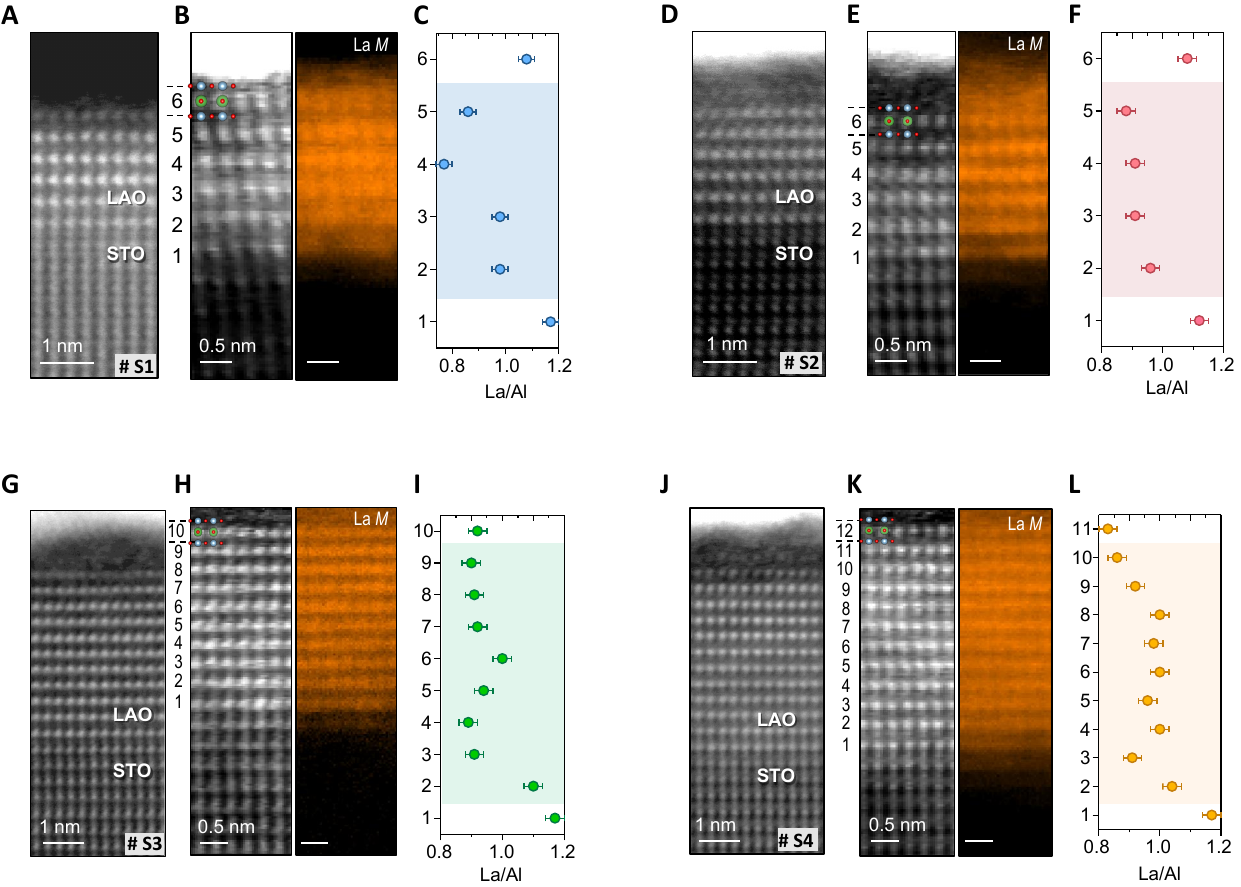}%
\end{center}
\vskip -0.5cm
\caption{\label{fig2} From left to right, HAADF atomic resolution image of LAO/STO interface, simultaneous ADF signal together with La elemental map and lateral averaged La/Al ratio of S1 (\textbf{a}, \textbf{b} and \textbf{c}), S2 (\textbf{d}, \textbf{e} and \textbf{f}), S3 (\textbf{g}, \textbf{h} and \textbf{i}) and S4 (\textbf{j}, \textbf{k} and \textbf{l}), respectively. The shaded region in \textbf{c}, \textbf{f}, \textbf{i}, \textbf{l} represents area for which the average value of La/Al ratio is estimated. The inset in \textbf{b,e,h, and k} shows the LaAlO$_3$ structure viewed along the [110] crystallographic direction, with La in green, Al in blue and O in red.}
\end{figure*}

\begin{figure*}[t]
 \begin{center}
\includegraphics [width=16cm]{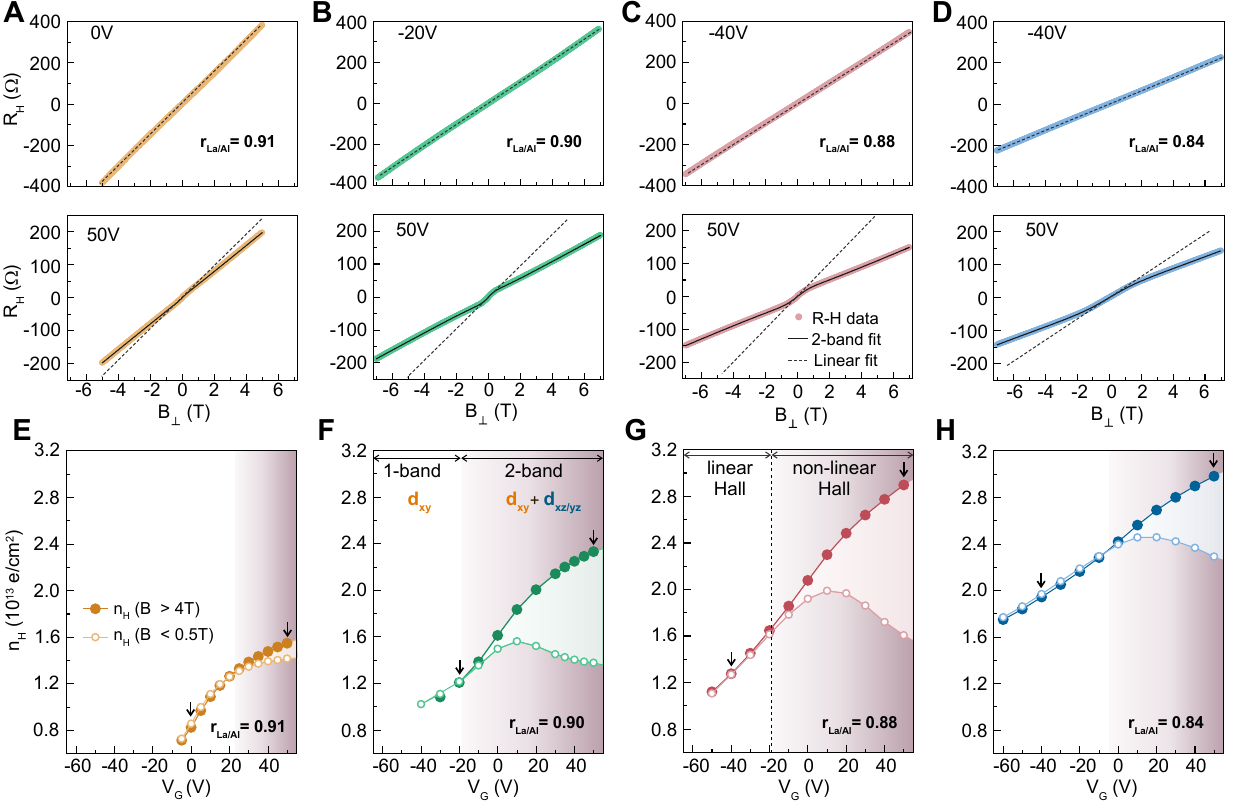}
\end{center}
\vskip -0.5cm
\caption{\label{fig3}  \textbf{a-d} The hall resistance as a function out-of-plane magnetic field $R_{H}(B_{\perp})$ for LAO/STO samples with r$_{La/Al}$ = 0.91, 0.90, 0.88, 0.84. The upper panel shows a curve for low gate voltage region of the phase diagram where $R_{H}(B_{\perp})$ shows a linear dependence, whereas, lower panel shows a curve at 50V with non-linear dependence in $R_{H}(B_{\perp})$. The black-dashed line indicates a linear fit of $R_{H}(B_{\perp})$. The black solid line in lower panel is a fit of two bands model (eq1 of supplementary note 4). \textbf{e-h} The evolution of carrier density as a function of V$_G$ extracted from the linear fit of $R_{H}(B_{\perp})$ for B$_{\perp}$ $<$ 0.5T (open circle) and B$_{\perp}$ $>$ 4T (solid circle). The shaded area is indicating the region of V$_G$ where $R_{H}(B_{\perp})$ is nonlinear i.e. two bands regime.  In the shaded region, n(B$_{\perp}$ $<$ 0.5T) vs. V$_G$ curve is not representing the true value of carrier density due to nonlinear $R_{H}(B_{\perp})$, whereas, n(B$_{\perp}$ $>$ 4T) is nearly equal to total carrier density $n_{T}=n_{d_{xy}}+n_{d_{xz/yz}}$. Two arrows are the selection of gate voltage for which $R_{H}(B_{\perp})$ curves are shown in the upper panels.}
\end{figure*}

\begin{figure*}[t]
\begin{center}
\includegraphics [width=16cm]{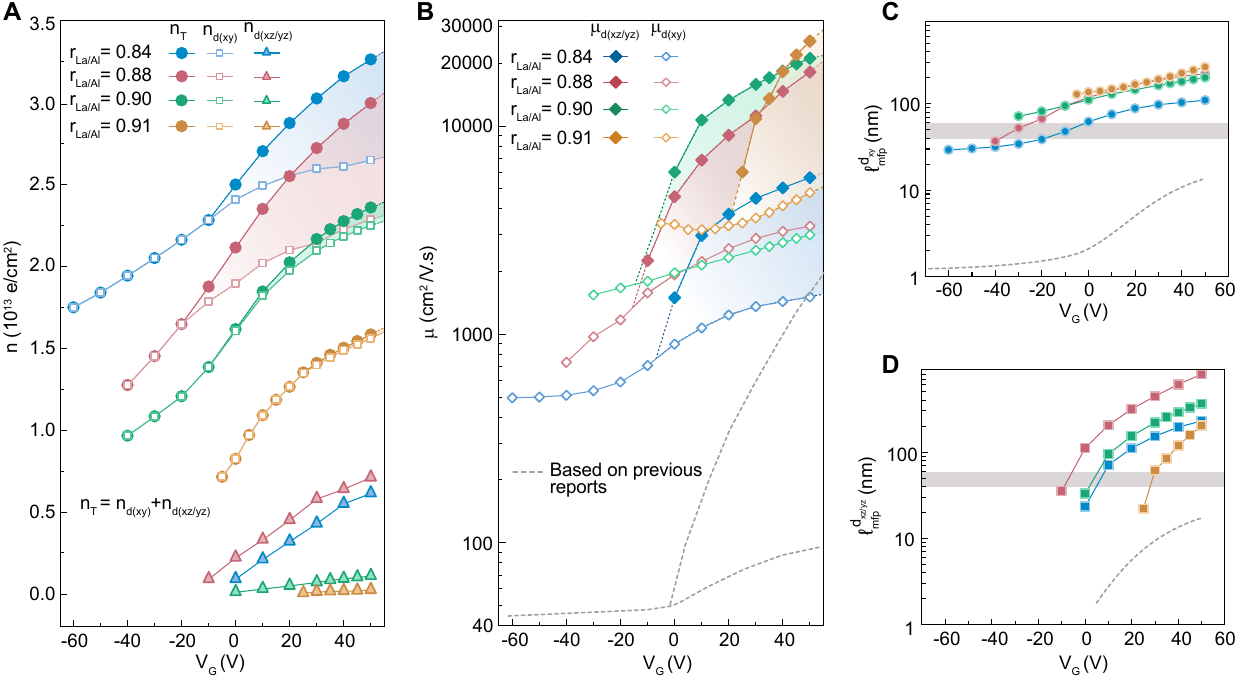}%
\end{center}
\vskip -0.5cm
\caption{\label{fig4} \textbf{a} The evolution of total carrier density n$_T$ (solid circle) and carrier density for electrons occupied in lower energy d$_{xy}$ band, n$_{d(xy)}$, (open square) and upper d$_{xz/yz}$ band, n$_{d(xz/yz)}$ (solid triangle), as a function of V$_G$, extracted from two-band model fit. \textbf{b} The extracted values of corresponding electron mobility $\mu_{d(xy)}$ and $\mu_{d(xz/yz)}$ as a function of V$_G$. The shaded area represents two-band region in the phase diagram. \textbf{c,d} The estimated value of mean free path for d$_{xy}$ electrons $l_{mfp}^{d_{xy}}$ and d$_{xz/yz}$ electrons $l_{mfp}^{d_{xz/yz}}$ respectively. The shaded gray area is the estimated value of coherence length adapted from Ref. \cite{Singh2018}.}
\end{figure*}

\vskip 0.4cm
On the other hand, the origin of the 2DEG in LAO/STO interfaces has been discussed in terms of either oxygen vacancies \cite{Herranz2007,Kalabukhov2007,Siemons2007} or polar discontinuity \cite{Nakagawa2006}. The former mechanism depends on external doping, while the latter mechanism is supported in view of the observation of conducting interfaces only beyond a critical thickness for LAO \cite{Liping2014}. Interestingly, a dependence of the interfacial conductivity on cation stoichiometry of the LAO layer was observed by Warusawithana et al. \cite{Warusawithana2013}, revealing that conductivity only appears in Al-rich LAO layers below a critical ratio La/Al $<$ 0.97. However, how the variation of the La/Al ratio affects fundamental parameters such as carrier density, mobility or the superconducting transition remains an open issue. In this work, we have investigated the transport properties of LAO/STO interfaces that show a consistent variation of La/Al ratio and remarkably large mobilities $\mu$ ranging between 5x10$^2$ - 3×10$^3$ cm$^2$/Vs, at low electrostatic doping (i.e. in the depletion region of the quantum well) and between 5x10$^3$ - 2x10$^4$ cm$^2$/Vs at high doping (in the accumulation regime). These values are more than an order of magnitude larger than previously reported on superconducting interfaces \cite{Herranz2015,Joshua2012,Biscaras2012,Singh2018,Singh2017,Nicola2019}. We demonstrate that a small variation of the La/Al ratio in Al-rich LaAlO$_3$, as identified in X-ray photoelectron spectroscopy (XPS) and scanning transmission electron microscopy coupled with electron energy-loss spectroscopy (STEM-EELS) measurements, provides an excellent degree of freedom, in addition to electrostatic doping, to consistently alter carrier densities, mobilities and, remarkably, the onset of superconductivity. In contrast to previous reports, where superconductivity is either absent or suppressed below the Lifshitz transition \cite{Herranz2015,Caviglia2008,Joshua2012,Biscaras2012,Singh2018,Singh2017,Nicola2019}, we observe a significant enhancement of T$_c$ for electrostatic doping below the Lifshitz transition. We attribute this result to the large mobility, which allows the mean free path (at low doping) to be comparable to the superconducting coherence length (i.e. $l_{mfp}$ $\sim$ $\xi$), which is key to achieving d$_{xy}$-superconductivity.                      
\vskip 0.5cm
\textbf{\large Film growth and characterization}\\
Thin LAO films were grown by pulsed laser deposition (PLD) on top of (001) oriented TiO$_2$-terminated STO substrates (see supplementary note 1 for growth process details). The epitaxial layer-by-layer growth of the films was confirmed by intensity oscillations of reflection high-energy electron diffraction (RHEED), as shown in Figure 1d, e. Films were investigated by Atomic Force Microscopy (Figure 1f), which showed a smooth surface with single unit cell steps. The film composition of La/Al ratio was investigated using X-ray photoemission spectroscopy (XPS). Here, we present results for two 5 unit cell (uc) (S1 and S2), one 10 uc thick (S3) and one 11 uc (S4) thick film. Figure 1g shows XPS spectra of Sr 3d, La 4d, Al 2s and Al 2p peaks for all samples. As expected, thicker films have much weaker intensity of substrate peaks (Ti and Sr). From the areas of La 4d and Al 2p peaks we estimated ratio between La and Al, assuming relative sensitivity factors of 6.52 and 0.54, correspondingly and with an error of about $\sim$ $\pm$1 $\%$ (see supplementary note 2). Due to charging effects, the analysis required an alignment of the core levels to the position of the Sr 3d peak. The results show that all films are La deficient (Figure 1i), in agreement with previously published results on films grown by PLD and MBE (see Figure 5g) \cite{Warusawithana2013}. The variation in the LAO film stoichiometry is likely due to the laser energy density during the film growth (supplementary figure 1). More details about the influence of growth parameters on the film composition are discussed in the Supporting Information. 

\vskip 0.4cm
Figure 2 shows the high-angle annular dark field (HAADF) atomic resolution images of LAO/STO heterostructures along with the La M elemental maps and the resulting averaged La/Al ratios for each atomic plane. These data indicate an inhomogeneous distribution of chemical composition, with a higher concentration of La near the interface in all samples, and at the uppermost plane of the 5 u.c thick LAO films of S1 and S2. The inhomogeneous distribution of La and Al near the interface can be caused by interdiffusion which is evident from atomic-resolution elemental maps of all elements including oxygen, see supplementary figure 2. In particular, the Ti L-edge elemental maps show that Ti is substituting Al within the first LAO layer in all samples. The colored regions of the averaged La/Al ratio panels (c, f, i and l) show the atomic planes having a moderate deviation of the La/Al ratio. The averaged La/Al ratios obtained from the HAADF images have slightly higher absolute values than those obtained from the XPS measurements, but follow the same trend as shown in figure 1i.

\vskip 0.4 cm
\textbf{\large Electrical transport in normal state}\\
The sheet resistance at T = 300 K decreases for smaller La/Al ratios (Figure 1h), indicating a correlation between electrical transport and stoichiometry. Since it is well-established that the electrical resistivity of the properly annealed crystalline LAO/STO interface does not depend on the thickness of the LAO overlayer \cite{Thiel2006, Herranz2012, Liu2013}, thus the variation in electrical resistivity in our samples can be attributed to the difference in the La/Al ratio. We emphasize that all samples show a metallic behavior in all the analyzed range of temperatures (2K-300K), with little sign of an increase in resistance at low temperatures (see supplementary figure 3), indicating the absence of weak localization effects \cite{Biscaras2010}. As discussed below, this agrees with the remarkably large values of the electronic mobility observed in all samples. In addition, carrier density is almost temperature independent in the interval 2K-300 K (see supplementary figure 4) in contrast to majority reports where a large reduction of carrier density was observed at low temperatures attributed to charge freeze out on charged crystal defects such as oxygen vacancies \cite{Huijben2013}. This again proves the high quality of interfaces studied in this work.
\vskip 0.4cm
We investigated the effect of the La/Al ratio (r$_{La/Al}$) on the evolution of carrier density at low temperature as a function of back gate voltages V$_G$. In the depleted regime, corresponding to negative or low positive V$_G$, we found a linear response of the transverse Hall resistance R$_{H}$ as a function of the applied out-of-plane magnetic field ($B_{\perp}$) in all samples (see top panels of Figures 3a-d). In contrast, in the overdoped regime, corresponding to large positive V$_G$, we observed non-linear R$_{H}$($B_{\perp}$), indicating that a second band is filled (bottom panels of Figures 3a-d) as the V$_G$ is increased. The transition from the linear to non-linear R$_{H}$($B_{\perp}$) is usually ascribed to a Lifshitz transition from d$_{xy}$ single-band to d$_{xy}$+d$_{xz/yz}$ two-band transport regimes when the d$_{xz/yz}$ subband is occupied at higher V$_G$ (see figure 1b, c for sketch of band structure). The transition can be better observed by obtaining the Hall carrier density $n_{H}$=$B_{\perp}$/eR$_{H}$ through linear fits of $R_{H}(B_{\perp})$ in the limits $B_{\perp}$ $<$ 0.5T and $B_{\perp}$ $>$ 4T, respectively (see figures 3e-h). The $n_{H}$ overlap at low V$_G$ in the linear regime (comparatively depleted region), in agreement with the single-band transport. In contrast, above the Lifshitz transition at larger V$_G$, the two values of the $n_{H}$ deviate from each other. 

\vskip 0.4cm
 In our samples the nonlinearity is observed at rather low magnetic fields $B_{\perp}$ $\le$ 2T, as can be seen in the bottom panels of Figures 3a-d. This is in contrast to previous reports that showed nonlinear behavior in $R_{H}(B_{\perp})$ extending for magnetic fields $B_{\perp}$ $>$ 5T \cite{Joshua2012,Biscaras2012,Singh2018,Singh2017}. We can attribute this to the distinct values of the electronic mobility in our samples, which are one order of magnitude larger than in previous reports. This is supported by the quantitative analysis of carrier densities and mobilities for each 3d-subbands using two-band model fit (see supplementary note 4).  

\begin{figure*}[t]
\begin{center}
\includegraphics [width=16cm]{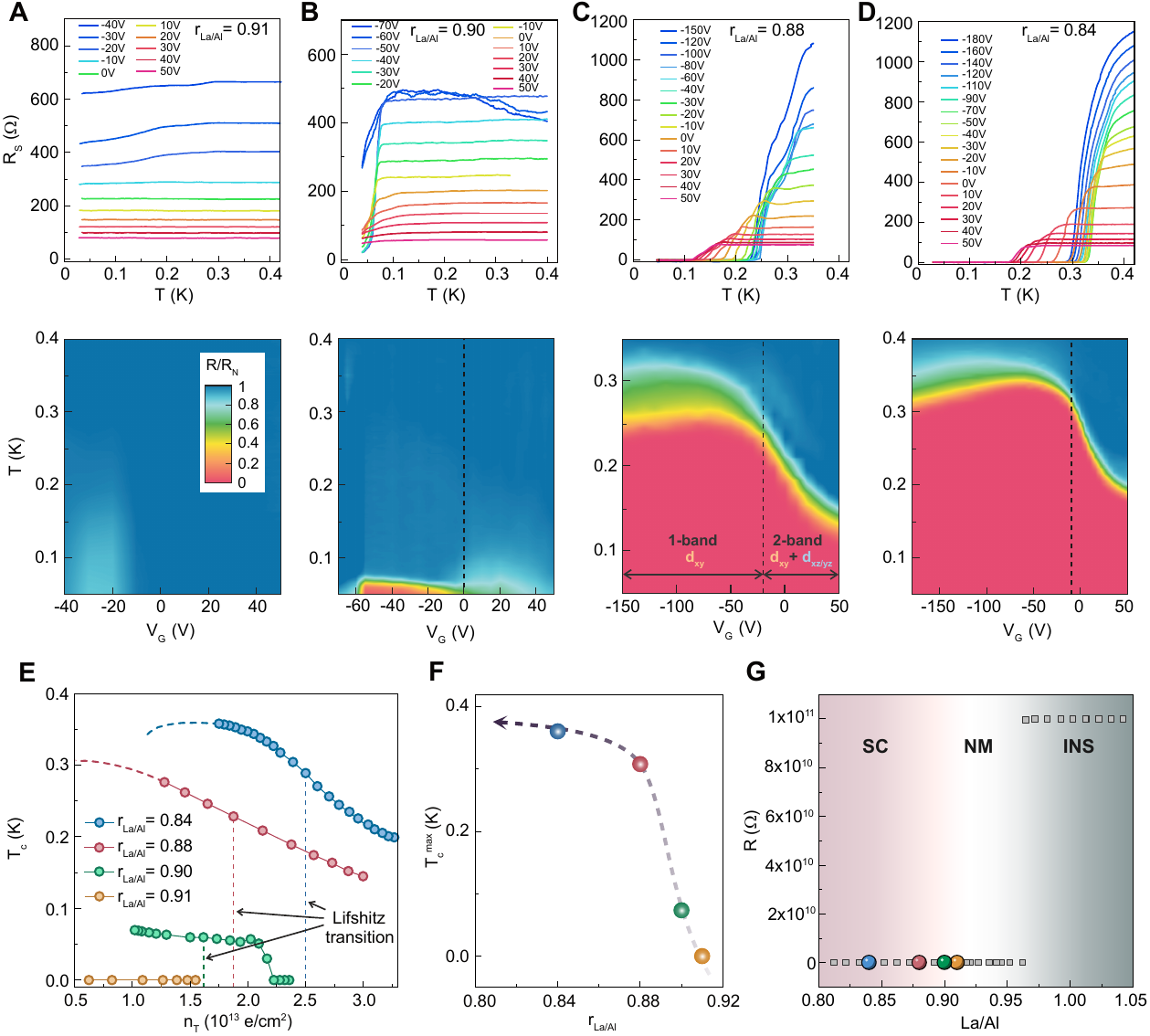}%
\end{center}
\vskip -0.5cm
\caption{\label{fig5} \textbf{a-d} Resistance vs temperature curves measured between 0.4K - 0.03K at different V$_G$ for samples with r$_{La/Al}$ = 0.91, 0.90, 0.88, 0.84. The lower panel shows the corresponding color plot of normalized resistance as a function of temperature and V$_G$. The dashed line in color plot represents V$_G$ for Lifshitz transition. \textbf{e} The modulation of critical temperature T$_c$ as a function of total carrier density n$_T$. Here the carrier density corresponding to the dash line is extracted from polynomial extrapolation of the data in Figure 3e-f. \textbf{f} The variation of maximum of superconducting $\mathrm{T}_{C}^{max}$ with La/Al ratio. \textbf{g} The conductivity in our samples is compared with the data from ref. \cite{Warusawithana2013}. The phase diagram is divided into three different regions, superconducting (SC) normal metal (NM) insulating (INS) based on the outcome of our result.}
\end{figure*}

\vskip 0.4cm
Figure 4a displays the V$_G$ dependence of the total carrier density n$_T$ and the carrier densities $n_{d_{xy}}$ and $n_{d_{xz/yz}}$ for electrons occupying $d_{xy}$ and $d_{xz/yz}$ subbands, respectively. Similarly, Figure 4b shows the data for the electronic mobilities $\mu_{d_{xy}}$ and $\mu_{d_{xz/yz}}$ as a function of V$_G$. In agreement with previous reports, we find that the majority of carriers are populating the lower energy d$_{xy}$ orbitals ($n_{d_{xy}}$ $\gg$ $n_{d_{xz/yz}}$) \cite{Joshua2012,Biscaras2012,Singh2018,Singh2017}. However, $n_{d_{xy}}$($\mu_{d_{xy}}$) and $n_{d_{xz/yz}}$($\mu_{d_{xz/yz}}$) display a systematic increase (decrease) with the reduction of the r$_{La/Al}$ ratio (figure 4a,4b). The values of $\mu_{d_{xy}}$ and $\mu_{d_{xz/yz}}$ obtained in our samples are about 15 times larger than in previous reports \cite{Joshua2012,Biscaras2012,Singh2018,Singh2017}. This can be attributed to the cleaner interfaces obtained in the present case. In addition, the values of the carrier density in the $d_{xz/yz}$ band are significantly larger than those reported previously. For instance, in our samples with r$_{La/Al}$ = 0.88 and 0.84, the fraction of carriers in $d_{xz/yz}$ bands exceeds 20$\%$, as compared to less than 2$\%$ observed in previous works \cite{Biscaras2012,Singh2018,Singh2017}. This implies that a small variation of the r$_{La/Al}$ ratio strongly affects carrier densities and mobilities in both bands, providing an additional degree of freedom to electrostatic gating, where the range of doping is limited by dielectric properties of SrTiO$_3$.
\vskip 0.5cm

\textbf{\large Superconducting properties}

To investigate superconductivity, we measured the temperature dependence of the sheet resistance $R_{S}(T)$ between 0.03 and 0.45 K for different V$_G$. Figures 5a-d show $R_{S}(T)$ curves at different V$_G$ (upper panel) and the color plot of normalized resistance as a function of T and V$_G$ (lower panel). It is evident that the r$_{La/Al}$ has also a strong effect on the formation of superconducting ground state. The maximum of superconducting critical temperature $\mathrm{T}_{C}^{max}$, extracted from T$_C$-V$_G$ phase diagram, increases with reducing r$_{La/Al}$, as shown in the Figure 5f. For r$_{La/Al}$ = 0.91, the interface does not display any evidence of superconducting transition for any V$_G$ (Figure 5a) despite being metallic down to the lowest temperature of 30 mK. A weak superconducting transition is observed for r$_{La/Al}$ = 0.90 with a $\mathrm{T}_{C}^{max}$ $\approx$ 0.08 K at V$_G$ $\approx$ -60 V (Figure 5b). In this sample, $R_{S}$ saturates to a residual resistance at low temperature, suggesting presence of non-percolating superconducting regions in the sample, as shown in supplementary figure 5 where superconducting regions embedded in metallic matrix are sketched. On the other hand, a clear zero-resistance state with $\mathrm{T}_{C}^{max}$ $\approx$ 0.25K at V$_G$ $<$ -50 V is achieved in the r$_{La/Al}$ = 0.88 sample (bottom panel in Figure 5c), where $R_{S}(T)$ displays a broad superconducting transition suggesting a competition between superconducting and metallic regions (supplementary figure 5). Interestingly, confirming a systematic trend, a sharp superconducting transition is observed in the samples with the lowest r$_{La/Al}$ = 0.84, with $\mathrm{T}_{C}^{max}$ $\approx$ $\approx$ 0.35K at gate voltages V$_G$ $\approx$ -50 V (bottom panel in Figure 5d). 
\vskip 2cm
\textbf{\large Discussion}\\
Our results provide important insights into the nature of superconductivity in the oxide q-2DEGs. According to the conventional Bardeen-Cooper-Schrieffer (BCS) scenario, pairing through $d_{xz/yz}$ rather than $d_{xy}$ bands is more favorable for superconductivity, due to the larger density of states of the former \cite{Bardeen1957}. In line with this, both experimental and theoretical studies show either the absence or strong suppression of superconductivity in single d$_{xy}$-band 2DEGs, in a region of the phase diagram where the electronic mobility lies below $\mu_{d_{xy}}$ $<$ 50 cm$^2$/V.s \cite{Herranz2015,Joshua2012,Biscaras2012,Singh2018,Singh2017,Nicola2019}. In striking contrast, the critical temperature T$_c$ of the samples analyzed in this work is not suppressed below the Lifshitz transition. As shown in the phase diagram (lower panel in Figure 5), for samples with r$_{La/Al}$ = 0.88 and 0.84 the T$_c$ increases substantially already below the Lifshitz transition, rising to $>$ 15$\%$ at maximum T$_c$, and further suppressed for higher doping. We attribute this remarkable observation to the large mobility of electrons in the $d_{xy}$ band, with values $\mu_{d_{xy}}$ $>$ 600 cm$^2$/V (see figure 4b), i.e., more than an order of magnitude larger than usually reported elsewhere. Strikingly, as inferred from data displayed in Figure 5e, the Lifshitz transition does not appear at a universal carrier density as suggested in previous observations \cite{Joshua2012}, indicating that the formation of the quantum well may evolve differently for different samples. These observations demonstrate that superconductivity at the LaAlO$_3$/SrTiO$_3$ interface can be achieved in the clean limit, opposite to previous reports where it was reported only in the dirty limit \cite{Singh2018, Singh2019}, where the mean free path is reduced to $l_{mfp}$ $<$ 1 nm in the depleted region. Our data (figures 4c, d) show that the $l_{mfp}$ in our samples is comparable to the $\xi$ = 40-60 nm. This observation is important for the search of theoretically predicted topological superconductivity in these systems, as it should minimize drastically the detrimental effect of disorder on non-conventional electron pairing \cite{Perroni2019, Lepori2021, Singh2022}. We also note that our results may allow studying pairing mechanism in single $d_{xy}$-band superconductivity in (001)-oriented LaAlO$_3$/SrTiO$_3$ interfaces in the dilute limit mediated by ferroelectric fluctuations through soft transverse optical phonons \cite{Edge2015,Shota2018,Shota2019,Maria2020,Gastiasoro,Yue,Pavel}. This is in line with recent observations of enhanced transition temperatures near the ferroelectric quantum critical point of SrTiO$_3$ \cite{Stucky2016,Rischau2017,Tomioka2019,Ahadi2019,Herrera2019}, and the possible co-existence of ferroelectricity and superconductivity \cite{Rischau2017,Russell2019}. Similar mechanisms have been proposed for the recently found superconductive KTaO$_3$ interfaces, in which pairing occurs in 5d instead of 3d orbitals \cite{Liu2021}. In particular, a soft optical mode induced by the inversion-breaking at the interface has been suggested for inter-orbital paring, a mechanism that could be general for quantum paraelectric interfaces \cite{Liu2023}. Finally, we also show that both electrical transport and superconductivity in the LaAlO$_3$/SrTiO$_3$ q-2DEG is systematically correlated with cation stoichiometry of the LAO film. While the exact mechanism of this effect cannot be elucidated from the existing data, it may be related to the intrinsic charge compensation by defects \cite{Warusawithana2013}. Remarkably, our sample with highest La/Al ratio (r$_{La/Al}$ = 0.91, S4) has striking similarity with previously reported SrCuO$_2$-capped q-2DEG showing temperature-independent carrier concentration and very high electron mobilities of up to $4\times 10^4$ cm$^2$/Vs, but no superconducting transition \cite{Huijben2013}. This suggests that the electrical transport in the LAO/STO interface is determined both by oxygen exchange kinetics and cation stoichiometry of the LAO film. 

\vskip 5cm
\textbf{\large Acknowledgements}\\
The authors thank Kyle Shen and Jak Chakhalian for useful discussions. We acknowledge financial support from Projects No. PID2020-118479RBI00 and Severo Ochoa FUNFUTURE (No. CEX2019-000917-S) of the Spanish Ministry of Science and Innovation (Grant No.MCIN/AEI/10.13039/501100011033) and by the Generalitat de Catalunya (2021 SGR 00445). We also acknowledge support from the Swedish infrastructure for micro- and nanofabrication - MyFab. G.S. acknowledges financial support from the Beatriu de Pinós Programme and the Ministry of Research and Universities of the Government of Catalonia, with research Grant No. 2019 BP 00207. N.B. acknowledge the funding received by the ANR PRC (QUANTOP) and by the QuantERA ERA-NET Cofund in Quantum Technologies (Grant Agreement N. 731473) implemented within the European Union's Horizon 2020 Program (QUANTOX). The ICN2 is funded by the CERCA program/Generalitat de Catalunya. The ICN2 is supported by the Severo Ochoa Centres of Excellence programme, Grant CEX2021-001214-S, funded by MCIN/AEI/10.13039.501100011033. R.G. $\&$ W.Z. acknowledge financial support from National Key RD Program of China (2018YFA0305800), and the Beijing Outstanding Young Scientist Program (BJJWZYJH01201914430039).

\vskip 20cm

{\large\textbf{Supplementary Note 1. Thin films growth}} 

Thin LAO films were grown on 5×5 mm$^2$ large TiO$_2$-terminated STO substrates by pulsed laser deposition (PLD). The laser spot area on the target was set to 2 mm$^2$. The laser fluence was varied by changing the energy, which was measured after the focusing lens. The laser beam optics was adjusted to provide a true image of the laser aperture on the target. All films were deposited at an oxygen pressure of $10^{-4}$ mbar and heater temperature of 800 $^o$C. The target-to-substrate distance was set to 50 mm. The laser repetition rate was 1 Hz, corresponding to the growth rate of 40-50 pulses per unit cell layer. After deposition, all samples were annealed to 600 $^o$C at the oxygen pressure of 300 mbar for one hour to eliminate oxygen vacancies that could be produced in the STO substrate during the deposition process. Pulsed laser deposition is a very complicated process involving various parameters and mechanisms. The resulting thin film stoichiometry depends mainly on the distribution of species in the plasma plume, which in turn is primarily governed by the laser fluence on the target, partial background gas pressure, and target-substrate distance. Other parameters like position of the substrate in the plume and substrate temperature may also affect the film composition. At the oxygen background pressure below 10$^{-3}$ mbar, early reports found that increasing laser energy density results in the decreased La/Al ratio (\cite{Breckenfeld2013,Park2020}). We observed a similar trend in our films, see Supplementary Figure 1a. However, there is a large spread of film composition as a function of laser fluence, which may be caused by small variations of sample position in the plume. We emphasize that the sheet resistance at T = 300 K systematically increases in samples with higher La/Al ratio, Supplementary Figure 1b. 
\vskip 0.4 cm
{\large\textbf{Supplementary Note 2. XPS:}} 

XPS measurements have been performed by using Al K$_\alpha$ X-ray source (E = 1486.6 eV) and Argus CU hemispherical electron spectrometer from Scienta Omicron AB, P.O. Box 15120, SE-750 15 Uppsala Sweden. Due to unavoidable charging effects, core level peaks in spectra for all samples have been aligned to the position of Sr 3d peak, see Figure 1g. La/Al ratios of all samples were estimated from analysis of La 4d and Al 2p peak areas using Scofield database for relative sensitivity factors (RSF) for La and Al of 6.52 and 0.537, correspondingly [J. H. Scofield, Lawrence Livermore Lab. Rept 1973, UCRL-51326]. Each value of La/Al ratio is an average of 3 consecutive measurements. The corresponding error was estimated to be ±1 $\%$.  
\vskip 0.4 cm
{\large\textbf{Supplementary Note 3. Scanning Transmission Electron Microscopy - Electron Energy-Loss spectroscopy (STEM-EELS):}}

STEM-EELS was performed along the [110] pseudocubic zone axis in an aberration-corrected Nion U-Hermes 100, operated at 60 kV, and equipped with Dectris ELA direct electron detector, which offers improved detective quantum efficiencies, narrower point spread functions and superior signal-to-noise ratios. The convergence semi-angle was 32 mrad, the probe current ~20 pA, and the collection semi-angle 75 mrad. The elemental maps were generated with Digital Micrograph’s EELS quantification tool. We used a reference EEL spectrum acquired from a control sample, a commercial LAO substrate. The background in the spectra was removed using a power law fit, followed by integrating the La M4,5 (830 eV) and Al K (1560 eV) edges intensity by using integration windows with energy range of 900-1000 eV and 1575-1690 eV, respectively, excluding the near-edge fine structure in both cases. The fitting parameters were adjusted to obtain a La/Al elemental quantification of the control sample. The La/Al ratio of LAO thin films was calculated using the same procedure, and to neglect the influence of thickness and plural scattering in the quantification of all spectra in this work (including the reference) were acquired in regions with the same sample thickness (20 nm).

\vskip 0.4 cm
{\large\textbf{Supplementary Note 4. Two-band model:}}

In the two-band regime analysis, the Hall resistance can be defined as
\begin{widetext}
\begin{equation}
R_{H}=\frac{B}{e}\frac{\frac{n_{d_{xy}}\mu_{d_{xy}}^2}{1+\mu_{d_{xy}}^2B^{2}}+\frac{n_{d_{xz/yz}}\mu_{d_{xz/yz}}^2}{1+\mu_{d_{xz/yz}}^2B^{2}}}{\left[\frac{n_{d_{xy}}\mu_{d_{xy}}}{1+\mu_{d_{xy}}^2B^{2}}+\frac{n_{d_{xz/yz}}\mu_{d_{xz/yz}}}{1+\mu_{d_{xz/yz}}^2B^{2}}\right]+\left[\frac{n_{d_{xy}}\mu_{d_{xy}}^2B}{1+\mu_{d_{xy}}^2B^{2}}+\frac{n_{d_{xz/yz}}\mu_{d_{xz/yz}}^2B}{1+\mu_{d_{xz/yz}}^2B^{2}}\right]^2}
\end{equation}
\end{widetext}
Where, $n_{d_{xy}}$, $n_{d_{xz/yz}}$ and $\mu_{d_{xy}}$, $\mu_{d_{xz/yz}}$ are carrier densities and mobilities related to $d_{xy}$ and $d_{xz/yz}$ bands which are obtained by fitting $R_{H}(B_{\perp})$ using constraints for total carrier density $n_{T}=n_{d_{xy}}+n_{d_{xz/yz}}$  and longitudinal resistance $\frac{1}{R_{S}}=e(n_{d_{xy}}\mu_{d_{xy}}+n_{d_{xz/yz}}\mu_{d_{xz/yz}})$.

\renewcommand{\figurename}{Supplementary Figure}
\setcounter{figure}{0}

\begin{figure*}[t]
\begin{center}
\vskip 0.5cm
\includegraphics [width=16cm]{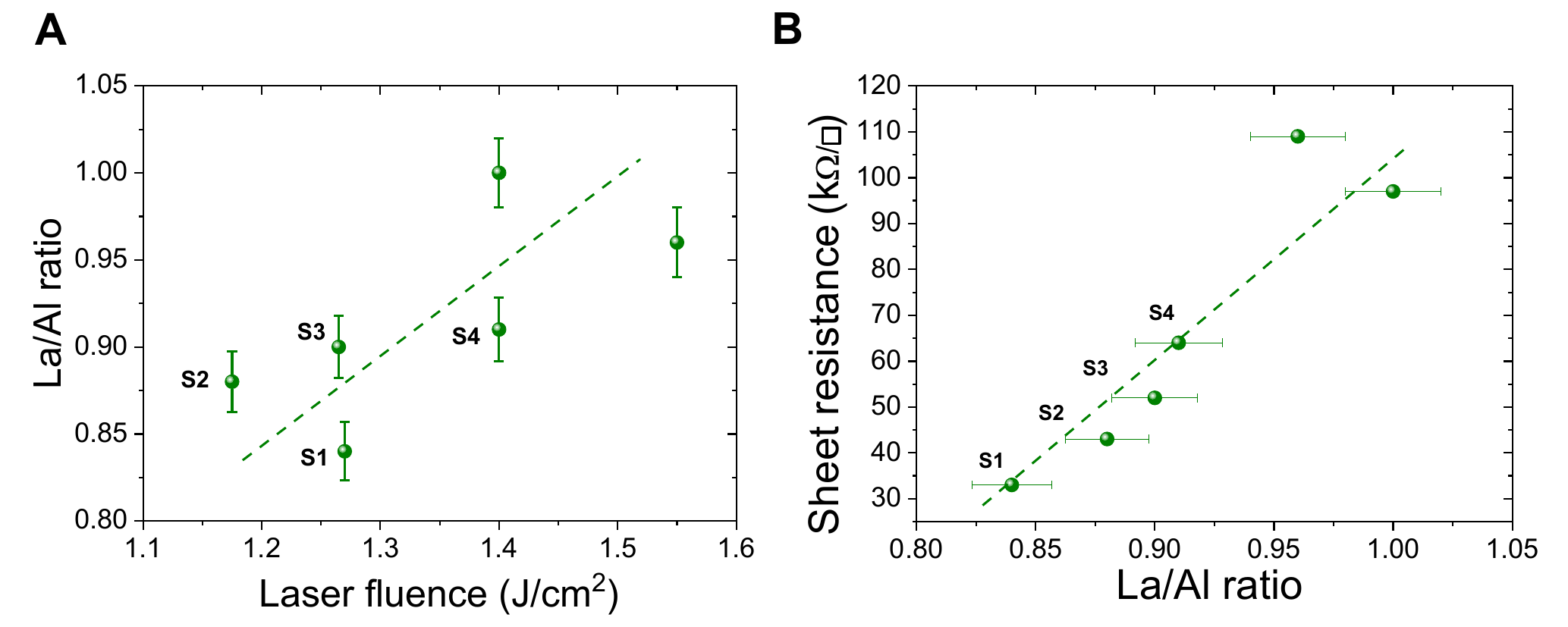}%
\end{center}
\vskip -0.5cm
\caption{\label{fig1} (a) Dependence of La/Al ratio obtained from XPS measurements on the laser fluence during pulsed laser deposition. (b) Room temperature sheet resistance of different LAO/STO samples as a function of La/Al ratio.}
\end{figure*}

\renewcommand{\figurename}{Supplementary Figure}

\begin{figure*}[t]
\begin{center}
\vskip 0.5cm
\includegraphics [width=14cm]{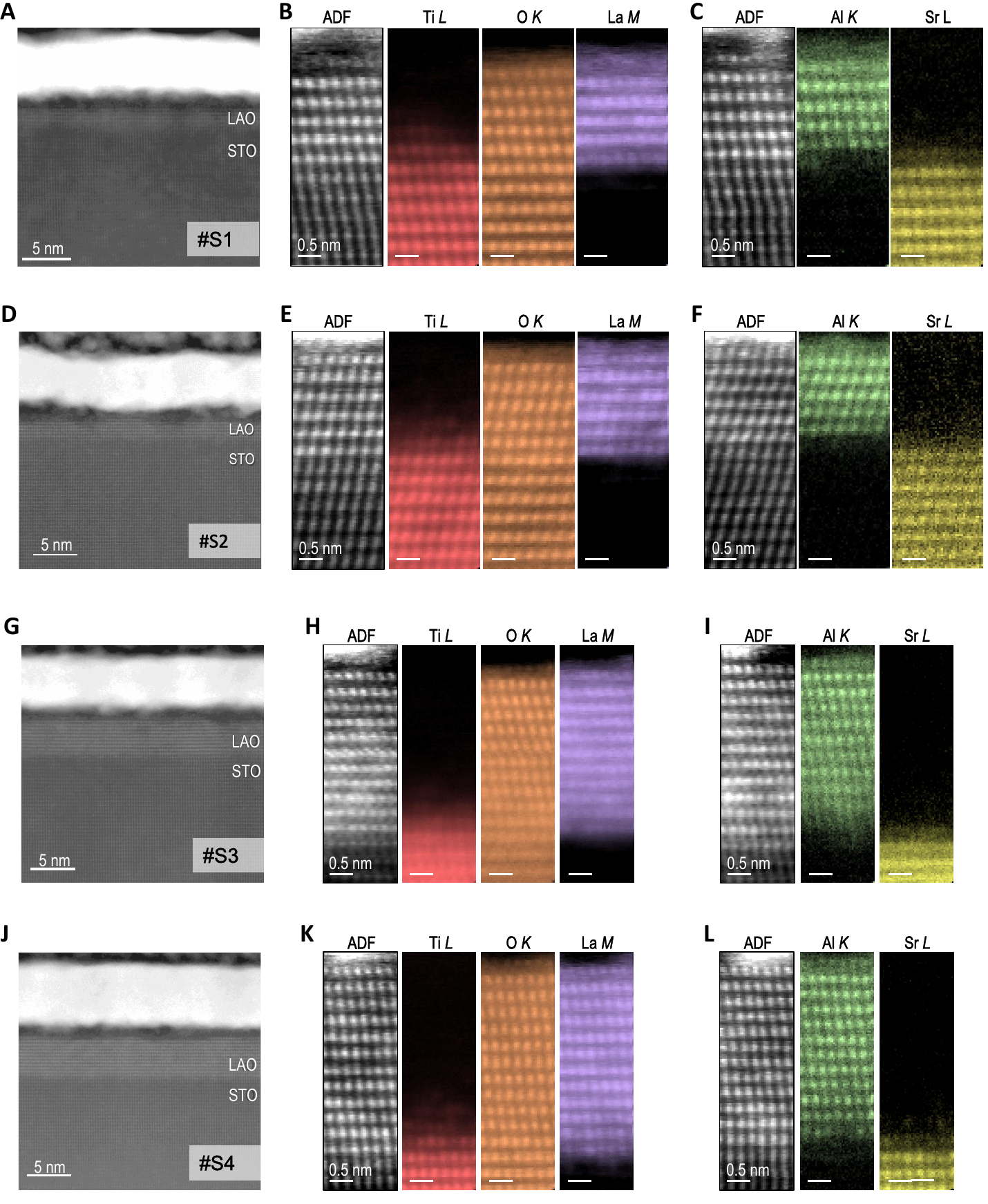}%
\end{center}
\vskip -0.5cm
\caption{\label{fig2} Atomic-resolution elemental maps of the rest of the cations using STEM-EELS. Since the energy range of the spectrometer is limited, two different sets of spectrum images were acquired to obtain the elemental maps of all cations of the heterostructure. Each row contains the atomic-resolution elemental maps of each sample of S1, S2, S3, and S4. The figures, from left to right, show the HAADF image (a,d,g,j) from where the spectrum images were acquired, the simultaneous ADF image of the spectrum used to obtain the Ti L (456 eV) edge, the O K edge (530 eV) and the La M (830 eV) edge elemental maps (b,e,h,k), and finally a second simultaneous ADF image of the spectrum image used to obtain the Al K-edge (1560 eV) and the Sr L-edge (1940 eV) elemental maps (c,f,i,l). }
\end{figure*}

\begin{figure*}[t]
\begin{center}
\vskip 0.5cm
\includegraphics [width=14cm]{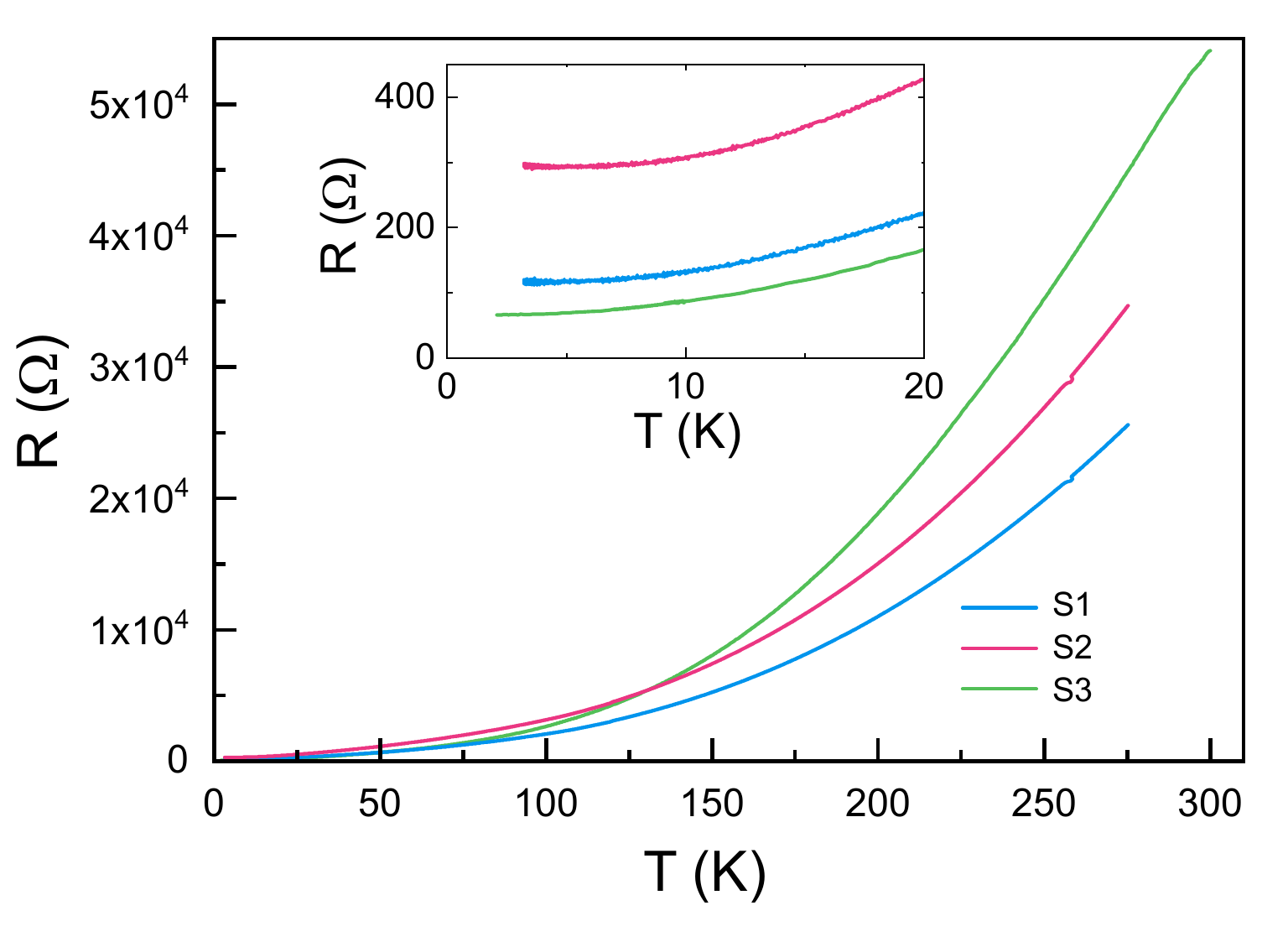}%
\end{center}
\vskip -0.5cm
\caption{\label{fig3} The sheet resistance as a function of temperature measured between 2-300K for samples S1, S2, S3. The inset shows a zoom below 20K.}
\end{figure*}

\begin{figure*}[t]
\begin{center}
\vskip 0.5cm
\includegraphics [width=15cm]{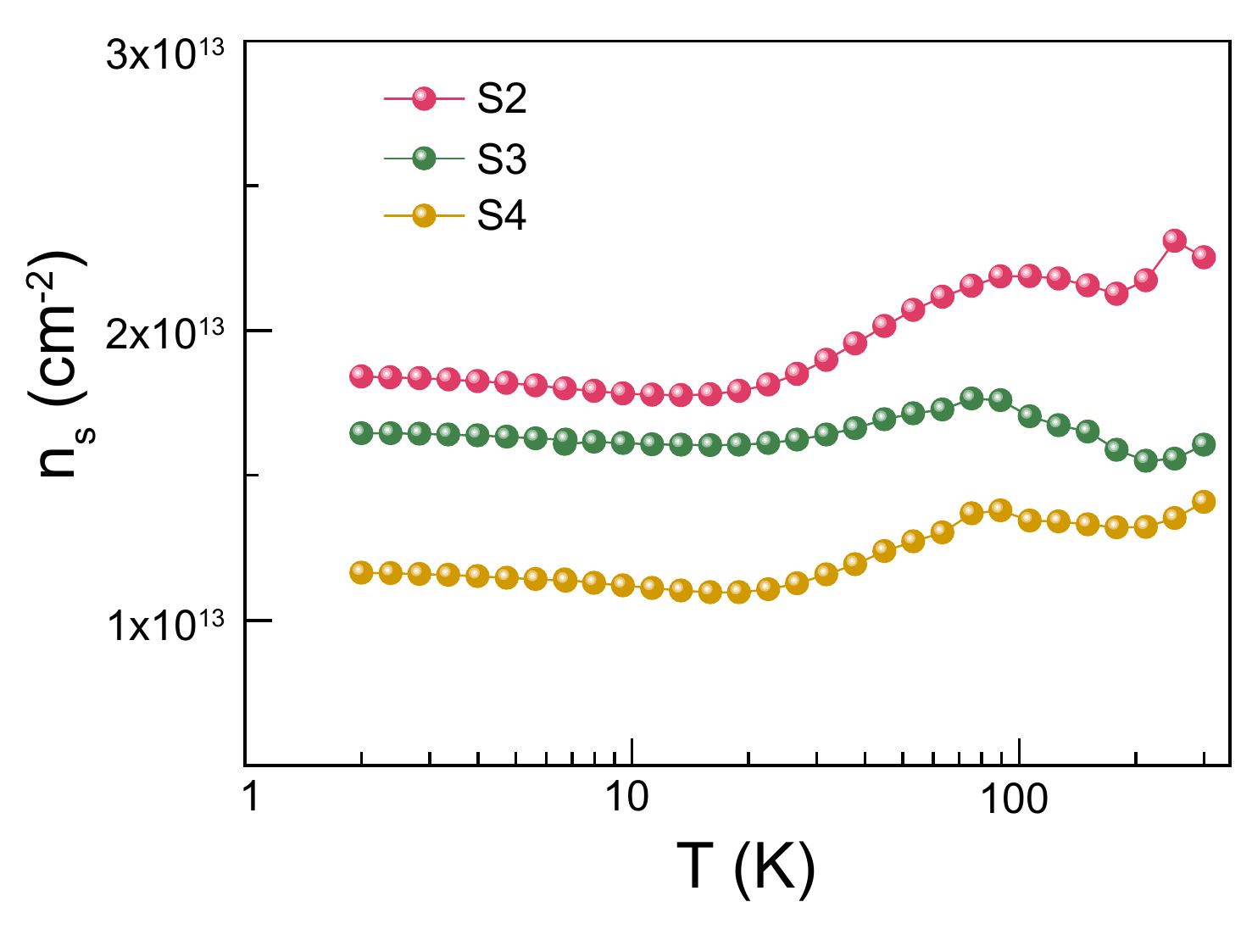}%
\end{center}
\vskip -0.5cm
\caption{\label{fig4} The carrier density n$_s$ as a function of temperature measured between 2K-300K for samples S2, S3 and S4.}
\end{figure*}

\begin{figure*}[t]
\begin{center}
\vskip 0.5cm
\includegraphics [width=16cm]{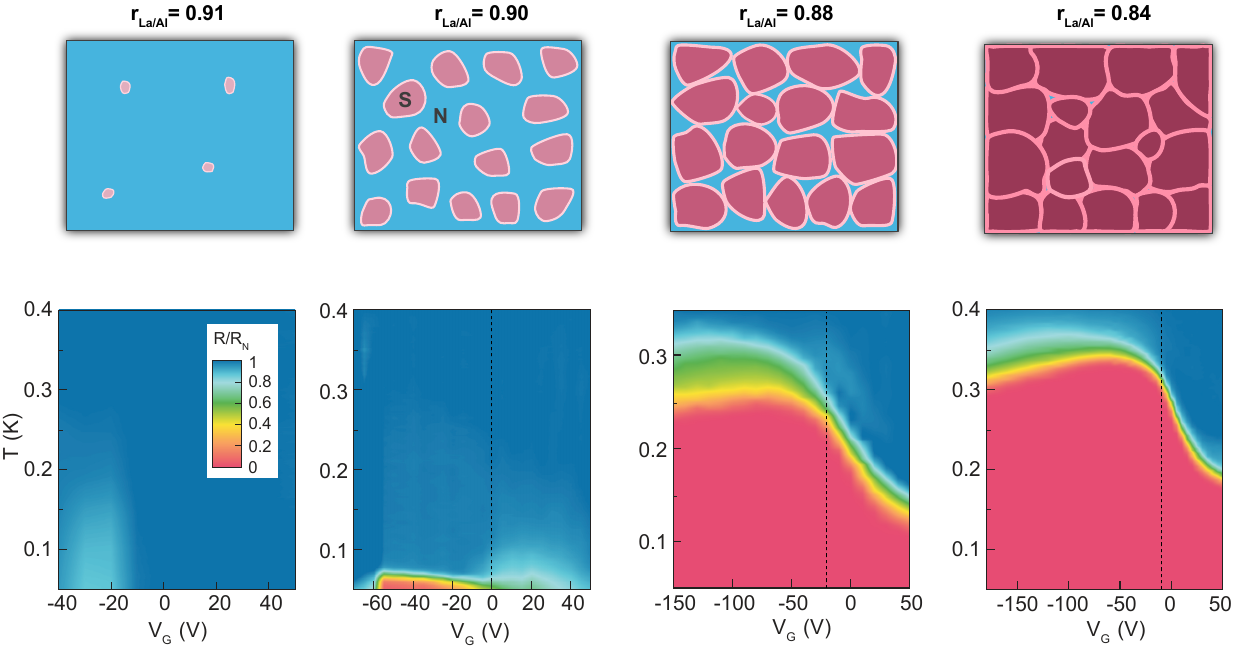}%
\end{center}
\vskip -0.5cm
\caption{\label{fig5} Upper panels show the sketch of superconducting puddles (in red) embedded in the weakly metallic matrix (in blue) for samples with r$_{La/Al}$ = 0.91, 0.9, 0.88, 0.84. In the lower panel, the corresponding measured superconducting phase diagram is presented.}
\end{figure*}

\end{document}